\documentclass{ieeeaccess}
\usepackage{cite}
\usepackage{amsmath,amssymb,amsfonts}
\usepackage{algorithmic}
\usepackage{graphicx}
\usepackage{textcomp}
\usepackage{array}
\newcolumntype{L}[1]{>{\raggedright\let\newline\\\arraybackslash\hspace{0pt}}m{#1}}
\newcolumntype{C}[1]{>{\centering\let\newline\\\arraybackslash\hspace{0pt}}m{#1}}
\newcolumntype{R}[1]{>{\raggedleft\let\newline\\\arraybackslash\hspace{0pt}}m{#1}}

\allowdisplaybreaks
\def\BibTeX{{\rm B\kern-.05em{\sc i\kern-.025em b}\kern-.08em
    T\kern-.1667em\lower.7ex\hbox{E}\kern-.125emX}}
\begin{document}
\history{Date of publication xxxx 00, 0000, date of current version xxxx 00, 0000.}
\doi{10.1109/ACCESS.2017.DOI}

\title{Adaptive Maximum Power Transfer for Movable device in Wireless Power Transfer system}

\author{\uppercase{Dong jun-Kim}\authorrefmark{1}}

\address[1]{50 Yonsei-ro, Seodamun-gu, Eng. C533, Seoul 03722, Republic of Korea  (e-mail: dongjungim20@yonsei.ac.kr)}


\markboth
{Author \headeretal: Preparation of Papers for IEEE TRANSACTIONS and JOURNALS}
{Author \headeretal: Preparation of Papers for IEEE TRANSACTIONS and JOURNALS}

\corresp{Corresponding author: Byung-Wook Min (e-mail: bmin@yonsei.ac.kr)}

\begin{abstract}
More and more applications are adopting the charging topology of wireless power transmission. However, most wireless charging systems can not charge mobile devices which are moving in position while charging. Currently, many commercialized wireless charging systems adopt an inductive coupling method, which has very short charging distances. In addition, the frequency of the two coupled coils that produce maximum power transfer keeps varying, depending on the coupling coefficient that relies on the separation between coils, and this tendency becomes more severe when the coupling is strengthened at a close charging distance by the phenomenon called frequency splitting. Therefore, the existing wireless power transmission system using a fixed operating frequency can't optimize power transmission for a fluctuating charging environment as the coupling between coils changes, and charging efficiency is greatly reduced by frequency splitting when charging at a very short distance. To solve this problem, we proposed the method of estimating the RX side power and mutual inductance using the information from the TX side such as input impedance rather than using a direct communication link which adds more cost and complexity. Also, we derived a mathematical model for the above estimation method. To prove this mathematical model, the proposed wireless power transmission system was implemented in a SIMULINK environment, and the system model was validated through simulation. Also comparison between the adaptive frequency tracking method and static impedance matching circuit is made by analyzing simulation results.
\end{abstract}

\begin{keywords}
: Wireless Power Transfer, Maximum Power Transfer, Movable Device, Gradient Descent, Mutual Inductance Estimation
\end{keywords}

\titlepgskip=-15pt
\maketitle
\section{Introduction}
\label{sec:introduction}
In modern society, as the number of mobile devices increases, wireless power transmission systems for charging mobile devices are also widely used. It is very comprehensive, from mobile devices such as mobile phones to home appliances and electric vehicles \cite{b1}-\cite{b6}. Currently, wireless charging systems for charging mobile devices use the ‘inductive coupling method’ to charge devices by placing them on the charging pad, supporting very short separation between transmitter and receiver. This method greatly hinders the portability of the device while charging, making it almost impossible to use the mobile device while moving, and in the end, it does not make much difference from the wired charging method. Wireless power transfer in the true sense must be able to charge the device while the device keeps moving. For this purpose, the most important thing is to transmit the maximum power regardless of the location of the receiver. The system proposed at \cite{b7},\cite{b17} transmits the power state of the RX side to the power transmitter(TX side) through a communication link like Bluetooth to control the system so that maximum power transmission becomes possible. However, this method improves the complexity, size, and price of the entire system. The communication delay caused by the communication link is a major factor hindering the real-time operation. In \cite{b8}, additional communication links can be removed by transmitting power and signal for communication through the same inductive link simultaneously. However, there is a problem that the efficiency for maximum power transmission is reduced as the carrier signal for power transmission is deformed by the process of modulation and demodulation for communication. In \cite{b9} and \cite{b10}, the current amplitude at the secondary side was predicted using the input impedance of the primary side. However, the distance between the two coils was set as a constant, so that the mutual inductance was some fixed value. In \cite{b11}, \cite{b12}, and \cite{b13}, both mutual inductance and load resistance were calculated, but they were not used to improve the system parameters like transmission efficiency. In \cite{b14} and \cite{b15}, coupling coefficient and load resistance were estimated through input impedance on the primary side, and the load power at the secondary side is controlled using the above information. In this study, the coupling coefficient was estimated through the phase angle of input impedance. At \cite{b16}, a gradient descent algorithm is used to find the frequency that produces omnidirectional wireless maximum power transfer. However, the gradient descent algorithm is applied to the primary side current, not a secondary side current. In this study, a wireless power transmission system was implemented to find the most efficient operating frequency for a given coupling environment to maximize transmitted power, and a gradient descent algorithm was applied to the power of the secondary side to find the optimized frequency in real-time. A novel method of estimating secondary side power using the magnitude of primary side input impedance without a direct communication link was studied.


\Figure[t!](topskip=0pt, botskip=0pt, midskip=0pt)[width=.45\textwidth]{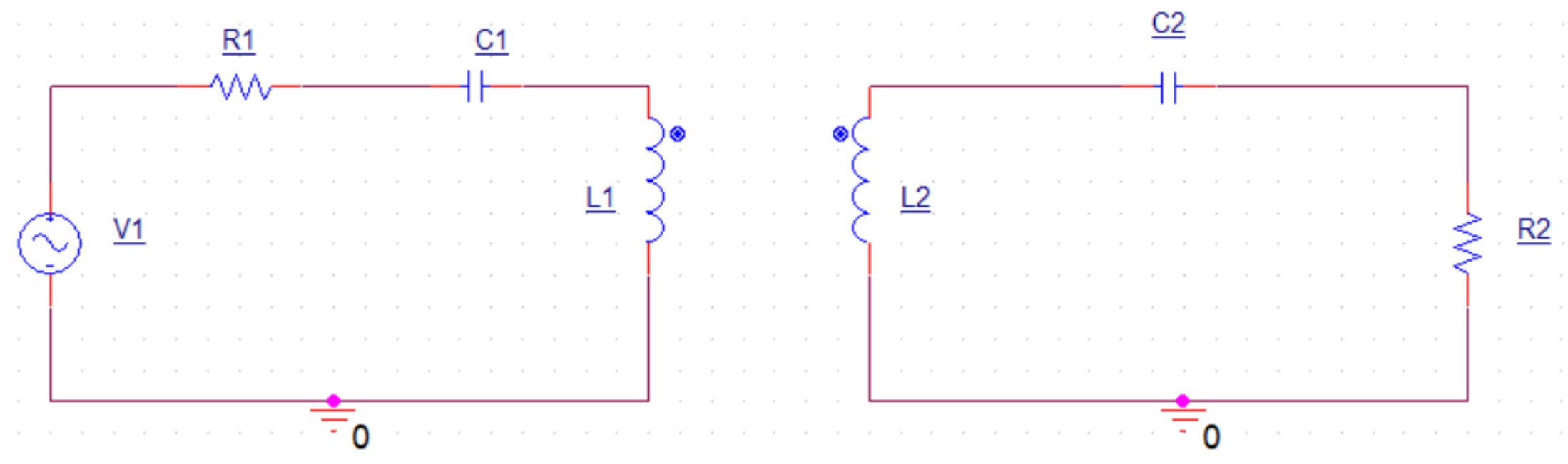}
{\textbf{Equivalent circuit of wireless power transfer system. L1 and L2 are coupled with mutual inductance of M.\label{circuit}}}

\section{Background}
\textbf{Wireless Power Transfer}\\
Various wireless power transmission systems have been proposed to date. Typically, magnetic induction, magnetic resonance, and electromagnetic wave-type wireless power transmission systems have been set as standards. The magnetic induction method and the magnetic resonance method use an electromagnetic induction phenomenon between coils. However, the magnetic resonance method is characterized by high energy transfer efficiency and long transmission distance, which are due to the same resonance frequency of the TX and RX coils, so prior technology research is actively underway.
In both the magnetic induction and magnetic resonance wireless power transmission system, two-side impedance matching in which the impedance of the power transmission side and the reception side are matched to the same value is essential to achieve high transmission efficiency. However, communication link is essential for two-side impedance matching in a movable device charging system where the mutual inductance between TX and RX coils can change in real-time, and matching both ends without feedback is extremely difficult. To solve this impedance matching problem, the dynamic frequency tracking method can be considered. However, it is crucial to recognize that in the case of using a dynamic frequency tracking scheme, it is highly probable to violate industrial, scientific, and medical frequency band, so there could be limitations in applying this method at industrial availability.

\section{Theoretical analysis}
\subsection{Equivalent circuit analysis}
First, we can think of a wireless power transfer equivalent circuit in Figure 1. In this case, $R_{2}$ can be expressed into $R_{Parasite} + R_{Load}$. When we solve the Kirchhoff equation at each loop, we can get two separate equations like below (1) and (2). Actually, we are driving a primary resonant circuit with a phase-shift half bridge which has DC supply voltage of $V_{D}$. If the harmonic components generated by the half-bridge inverter can be ignored, Fundamental Harmonic Approximation (FHA) could be applied, so V becomes the fundamental component of the half-bridge output voltage, $V_{1}(t) = \frac{2V_{D}}{\pi}sin(wt+\phi)$ where $V_{1} = V$.
\begin{flalign} 
    V &= I_{1}(R_{1}+JWL_{1}+\frac{1}{jwC_{1}})-jwMI_{2}\\
    0 &= -jwMI_{1}+I_{2}(R_{2}+jwL_{2}+\frac{1}{jwC_{2}})&&
\end{flalign}
By equating the above two equations, $I_{1}$ can be expressed like below.
\begin{align*} 
    V & =I_{1}(R_{1}+jwL_{1}+\frac{1}{jwC_{1}})-jwM\frac{jwMI_{1}}{R_{2}+jwL_{2}+\frac{1}{jwC_{2}}}\\
    & =I_{1}(R_{1}+jwL_{1}+\frac{1}{jwC_{1}}) + \frac{(wM)^{2}I_{1}}{R_{2}+jwL_{2}+\frac{1}{jwC_{2}}}\\
    & = I_{1}(R_{1}+jwL_{1}+\frac{1}{jwC_{1}}+\frac{(wM)^2}{R_{2}+jwL_{2}+\frac{1}{jwC_{2}}})
\end{align*}
Express the above equation for $I_{1}$.
\begin{flalign*} 
     I_{1} = \frac{V}{R_{1}+jwL_{1}+\frac{1}{jwC_{1}}+\frac{(wM)^2}{R_{2}+jwL_{2}+\frac{1}{jwC_{2}}}}&& 
\end{flalign*}
By substituting the above equation at (2), we can get the expression of $I_{2}$ like below.
{
\small
\begin{align*} 
    I_{2} &= \frac{jwMI_{2}}{R_{2}+jwL_{2}+\frac{1}{jwC_{2}}}\\
    &= \frac{jwM}{R_{2}+jwL_{2}+\frac{1}{jwC_{2}}}\frac{V}{R_{1}+jwL_{1}+\frac{1}{jwC_{1}}+\frac{(wM)^2}{R_{2}+jwL_{2}+\frac{1}{jwC_{2}}}}
\end{align*}
}%
By taking absolute value at each side of above equation, we can get the amplitude of $I_{2}$.
{
\small
    \begin{flalign*} 
        |I_{2}| &= \frac{|wM|}{|R_{2}+jwL_{2}+\frac{1}{jwC_{2}}|}\frac{|V|}{|R_{1}+jwL_{1}+\frac{1}{jwC_{1}}+\frac{(wM)^2}{R_{2}+jwL_{2}+\frac{1}{jwC_{2}}}|}\\
        &= \frac{wMV}{|R_{2}+j(wL_{2}-\frac{1}{wC_{2}})||R_{1}+j(wL_{1}-\frac{1}{WC_{1}})+\frac{(wM)^2}{R_{2}+j(wL_{2}-\frac{1}{wC_{2}})}|}&&
    \end{flalign*} 
}
Now, let $X_{1}$ = $wL_{1}-\frac{1}{wC_{1}}$ and $X_{2}$ = $wL_{2}-\frac{1}{wC_{2}}$ for simplicity. Then,
\begin{align*} 
   |I_{2}| &= \frac{wMV}{|R_{2}+jX_{2}||R_{1}+jX_{1}+\frac{(wM)^2}{R_{2}+jX_{2}}|}\\
   &=\frac{wMV}{|R_{2}+jX_{2}||R_{1}+jX_{1}+\frac{(wM)^2R_{2}-j(wM)^2X_{2}}{R_{2}^2+X_{2}^2}|}\\
   &=\frac{wMV}{|R_{2}+jX_{2}||R_{1}+\frac{(wM)^2R_{2}}{R_{2}^2+X_{2}^2}+j(X_{1}-\frac{(wM)^2X_{2}}{R_{2}^2+X_{2}^2})|}\\
   &= \frac{wMV}{\sqrt{R_{2}^2+X_{2}^2}\sqrt{(R_{1}+\frac{(wM)^2R_{2}}{R_{2}^2+X_{2}^2})^2+(X_{1}-\frac{(wM)^2X_{2}}{R_{2}^2+X_{2}^2})^2}}
\end{align*} 

Using MATLAB, |$I_{2}$| is plotted as a function of k and w where w = 2$\cdot\pi\cdot$f, and the resultant graph can be shown in Figure 2. All other element value is fixed according to the corresponding value of Table 1. We can show that as the coupling coefficient get bigger, local maxima are observed at two different frequency simultaneously, which is called ‘frequency splitting’ or ‘pole splitting’. Even before the frequency splitting, the frequency of maximum power transfer points keeps varying, according to the changes of coupling coefficient. These points are definitely hard things to handle in a conventional wireless power transfer system.

\begin{figure}[!t]
    \centering
    \includegraphics[width=.45\textwidth]{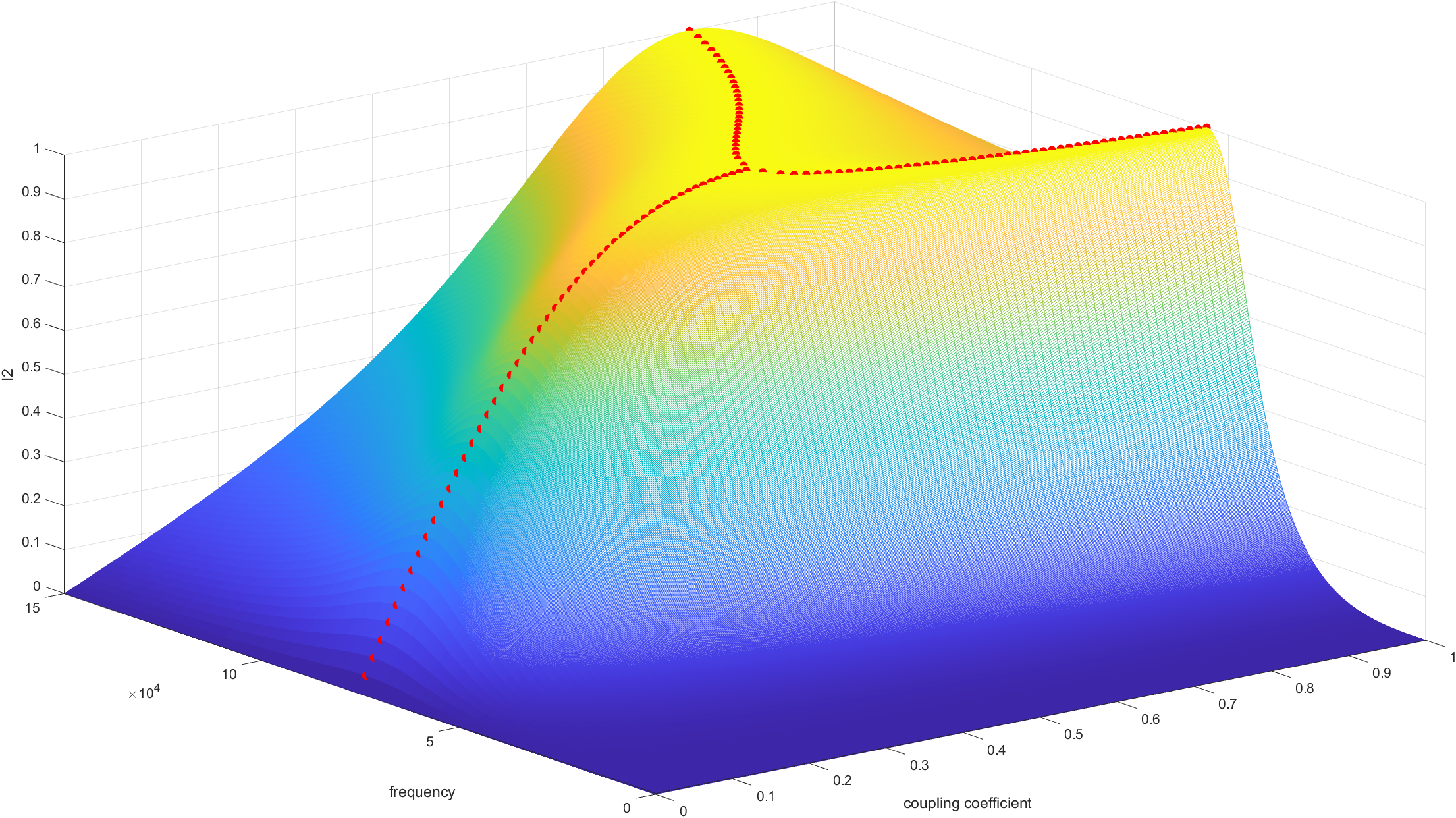}
    \caption{\textbf{3-D |$I_{2}$| Graph, with respect to f(operating frequency) and k(coupling coefficient). Red dots represent the local maximum points where the partial derivative in the k direction becomes zero, showing the prominent frequency splitting phenomenon.}}
    \label{fig:freq_split_graph}
\end{figure}

\subsection{Mutual inductance estimation }
To calculate $I_{2}$ from primary side information like $I_{1}$ and $V_{1}$, we need to estimate mutual inductance M also.
This can be derived from primary side input impedance $Z_{IN}$. Let’s think of the below equation.
\begin{flalign*} 
     I_{1} &= \frac{V}{R_{1}+jwL_{1}+\frac{1}{jwC_{1}}+\frac{(wM)^2}{R_{2}+jwL_{2}+\frac{1}{jwC_{2}}}}\\
     &= \frac{V}{Z_{IN}}&&
\end{flalign*}

We can calculate the mutual inductance between coils in two ways.
By (1) the magnitude of input impedance and (2) the phase angle of input impedance.\\
\textbf{Method1}: Using the magnitude of input impedance
\begin{flalign*} 
    Z_{IN} &= R_{1}+jwL_{1}+\frac{1}{jwC_{1}}+\frac{(wM)^2}{R_{2}+jwL_{2}+\frac{1}{jwC_{2}}}\\
    |Z_{IN}| &= \sqrt{(R_{1}+\frac{w_{2}M_{2}R_{2}}{R_{2}^2+X_{2}^2})^2+(X_{1}-\frac{w_{2}M_{2}X_{2}}{R_{2}^2+X_{2}^2})^2}&&
\end{flalign*}
Let's substitute A = $\frac{w^2R_{2}}{R_{2}^2+X_{2}^2}$, B = $\frac{w^2X_{2}}{R_{2}^2+X_{2}^2}$.
\begin{align*}
    |Z_{IN}|^2 & = (R_{1}+\frac{w_{2}MR_{2}}{R_{2}^2+X_{2}^2})^2+(X_{1}-\frac{w_{2}MX_{2}}{R_{2}^2+X_{2}^2})^2\\
   & = (R_{1}+M^2A)^2+(X_{1}-M^2B)^2\\
   & = R_{1}^2+2R_{1}AM^2+M^4A^2+X_{1}^2-2X_{1}M^2B+M^4B^2
\end{align*}
Let's simplify the above equation for M.\\
\\
{\small  $M^4(A^2+B^2)+2M^2(R_{1}A-X_{1}B)+R_{1}^2+X_{1}^2-|Z_{IN}|^2=0$}\\
\\
Let's substitute $\alpha =A^2+B^2$, $\beta = R_{1}A-X_{1}B$, $\gamma = R_{1}^2+X_{1}^2-|Z_{IN}|^2$ and $M^2$ = t( $>0$ ). Then we can derive a second-order equation and can apply the quadratic formula here easily like below.
\begin{align*} 
    \alpha t^2+2\beta t + \gamma = 0\\
    t = \frac{-\beta \pm \sqrt{\beta^2-\alpha\gamma}}{\alpha}\\
    M = \sqrt{\frac{-\beta \pm \sqrt{\beta^2-\alpha \gamma}}{\alpha}}
\end{align*}
Where,
\begin{align*} 
    \alpha &= (\frac{w^2R_{2}}{R_{2}^2+X_{2}^2})^2+(\frac{w^2X_{2}}{R_{2}^2+X_{2}^2})^2\\
    \beta &= R_{1}\frac{w^2R_{2}}{R_{2}^2+X_{2}^2} - X_{1}\frac{w^2X_{2}}{R_{2}^2+X_{2}^2}\\
    \gamma &= R_{1}^2+X_{1}^2-|Z_{IN}|^2\\
    X_{1} &= wL_{1}-\frac{1}{wC_{1}}\\
    X_{2} &= wL_{2}-\frac{1}{wC_{2}}
\end{align*}
\textbf{Method2}: Using the phase angle of input impedance
\begin{flalign*} 
    Z_{IN} &= R_{1}+jwL_{1}+\frac{1}{jwC_{1}}+\frac{(wM)^2}{R_{2}+jwL_{2}+\frac{1}{jwC_{2}}}\\
    \angle Z_{IN} &= \arctan \frac{X_{1}R_{2}^2+X_{1}X_{2}^2-w^2M^2X_{2}}{R_{1}R_{2}^2+R_{1}X_{2}^2+w^2M^2R_{2}}\\
    \tan{\angle Z_{IN}} &= \frac{X_{1}R_{2}^2+X_{1}X_{2}^2-w^2M^2X_{2}}{R_{1}R_{2}^2+R_{1}X_{2}^2+w^2M^2R_{2}}&&
\end{flalign*}
Let’s simplify the above equation for M.
{\normalsize
    \begin{eqnarray*} 
        M^2(\tan{\angle Z_{IN}} \cdot R_{2}w^2+w^2X_{2}) =\\ X_{1}R_{2}^2+X_{1}X_{2}^2-\tan{\angle Z_{IN}} \cdot (R_{1}R_{2}^2+R_{1}X_{2}^2)
    \end{eqnarray*}
}%
By solving the above simple quadratic polynomial equation, we can get a solution easily like below.
\begin{eqnarray*}
    M &=& \frac{1}{w}\sqrt{\frac{X_{1}R_{2}^2+X_{1}X_{2}^2-\tan{\angle Z_{IN}} \cdot (R_{1}R_{2}^2+R_{1}X_{2}^2)}{\tan{\angle Z_{IN}} \cdot R_{2}w^2+w^2X_{2}}}
\end{eqnarray*}
In my research, I choose \textbf{Method1} for estimating the secondary side current amplitude $|I_{2}|$.
Finally, the $|I_{2}|$ can be expressed like below.
\begin{eqnarray*} 
    |I_{2}| &= \frac{wV}{\sqrt{R_{2}^2+X_{2}^2}\sqrt{(R_{1}+\frac{(wM)^2R_{2}}{R_{2}^2+X_{2}^2})^2+(X_{1}-\frac{(wM)^2X_{2}}{R_{2}^2+X_{2}^2})^2}} \sqrt{\frac{-\beta \pm \sqrt{\beta^2-\alpha \gamma}}{\alpha}}
\end{eqnarray*}
Where,
\begin{align*} 
    \alpha &= (\frac{w^2R_{2}}{R_{2}^2+X_{2}^2})^2+(\frac{w^2X_{2}}{R_{2}^2+X_{2}^2})^2\\
    \beta &= R_{1}\frac{w^2R_{2}}{R_{2}^2+X_{2}^2} - X_{1}\frac{w^2X_{2}}{R_{2}^2+X_{2}^2}\\
    \gamma &= R_{1}^2+X_{1}^2-|Z_{IN}|^2\\
    X_{1} &= wL_{1}-\frac{1}{wC_{1}}\\
    X_{2} &= wL_{2}-\frac{1}{wC_{2}}
\end{align*}

\subsection{Gradient ascent algorithm}
Through the above discussion, we can estimate the amount of power delivered to the secondary side, when we know the operating frequency and exact value of each element consisting of the entire circuit. So, we can think of the relation between the operating frequency and RX side power at a fixed coupling coefficient(k) – which means fixed separation between coils. When we assume that the relation could be modeled like Figure 3's graph, we can apply a gradient descent algorithm to track the frequency at which the delivered power at the RX side is maxima. Considering the equation of $I_{2}$ and its 3-D graph for coupling coefficient and frequency, the previous assumption is quite valid without loss of generality.
Mathematically, the gradient ascent algorithm (which is exactly the inverse of the gradient descent algorithm) can be expressed below Figure 3.

\begin{figure}[ht]
    \centering
    \includegraphics[width=.45\textwidth]{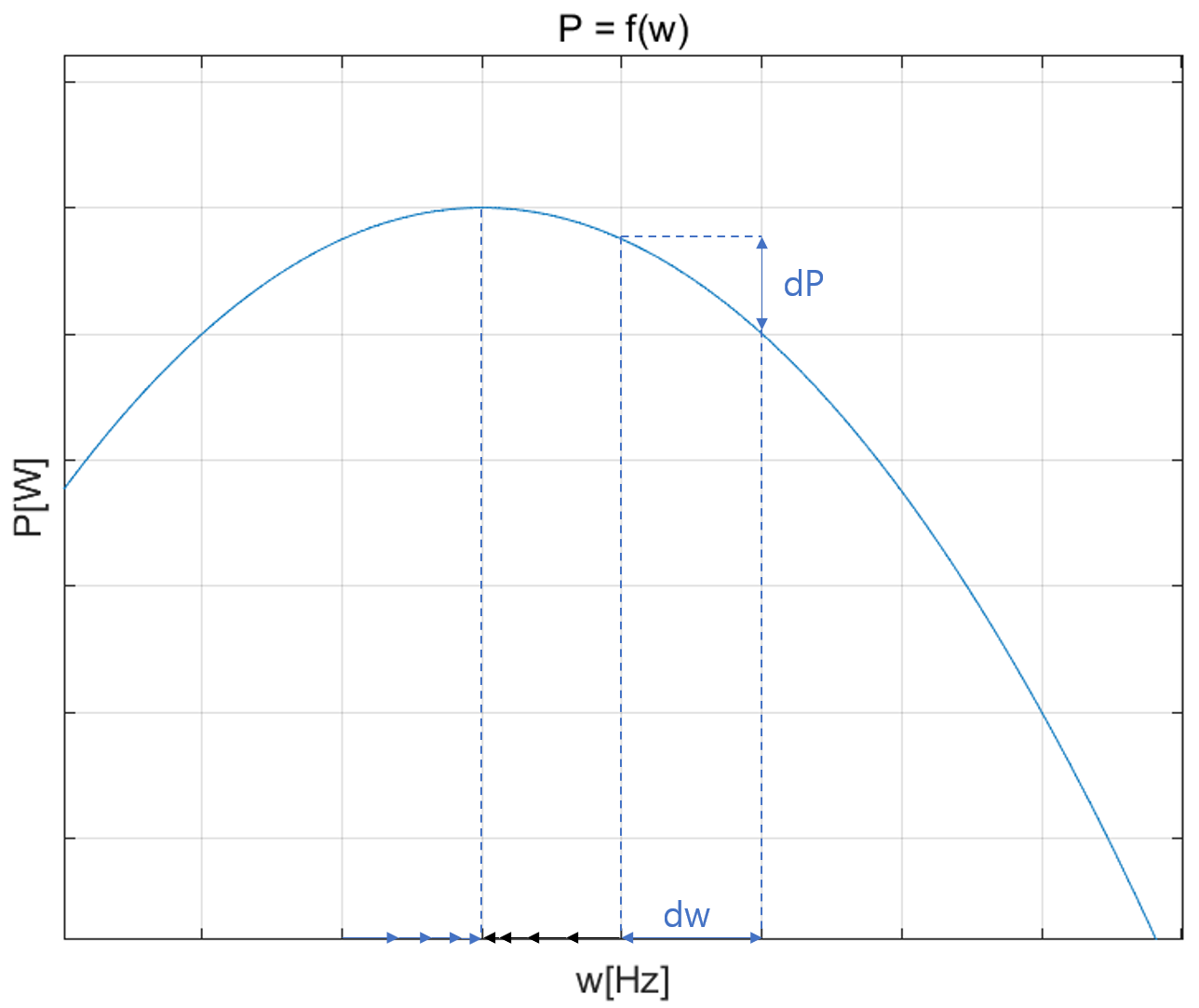}
    \caption{\textbf{Gradient Ascent Algorithm}}
    \label{fig:gaa}
\end{figure}
\begin{eqnarray*}
    \frac{dw}{dt} = +\alpha\frac{dP}{dw}
\end{eqnarray*}

The above equation describes the dynamics of an algorithm at continuous domain. we need to convert it into a discrete domain because we use digital controller like DSP or MCU to control the operating frequency of the system.

\begin{eqnarray*}
    w_{n} = w_{n-1} + \alpha(P_{n-1}-P_{n-2})\frac{w_{n-1}-w_{n-2}}{|w_{n-1}-w_{n-2}|}
\end{eqnarray*}

, where $\alpha$ is the learning coefficient which determines the speed of convergence and stability of the system. 

\section{Modeling methodology}
\subsection{Equivalent circuit analysis by SPICE}

Using the SPICE program, I modeled the resonant circuit part of the wireless power transfer system. Each side has the same resonant frequency of 72Khz. Considering the same scheme as Figure 1, each element's value is specified in Table 2.

\begin{table}[ht]
\begin{tabular}{|l|l|}
\hline
\textbf{Element} & \textit{\textbf{Value}} \\ \hline
V1               & \textit{10V}      \\ \hline
R1               & \textit{5ohm}              \\ \hline
R2               & \textit{5ohm}              \\ \hline
C1               & \textit{240.7nF}        \\ \hline
C2               & \textit{240.7nF}        \\ \hline
L1               & \textit{20uH}           \\ \hline
L2               & \textit{20uH}           \\ \hline
k                & \textit{0.1\textasciitilde  0.9}            \\ \hline
\end{tabular}
\caption{\textbf{Element value consisting of the circuit in Figure 1.}}
\label{Table:element_value}
\end{table}

After conducting an AC sweep simulation at various coupling coefficients, I got the below results in Figure 4.
\begin{figure}[ht]
    \centering
    \includegraphics[width=.5\textwidth]{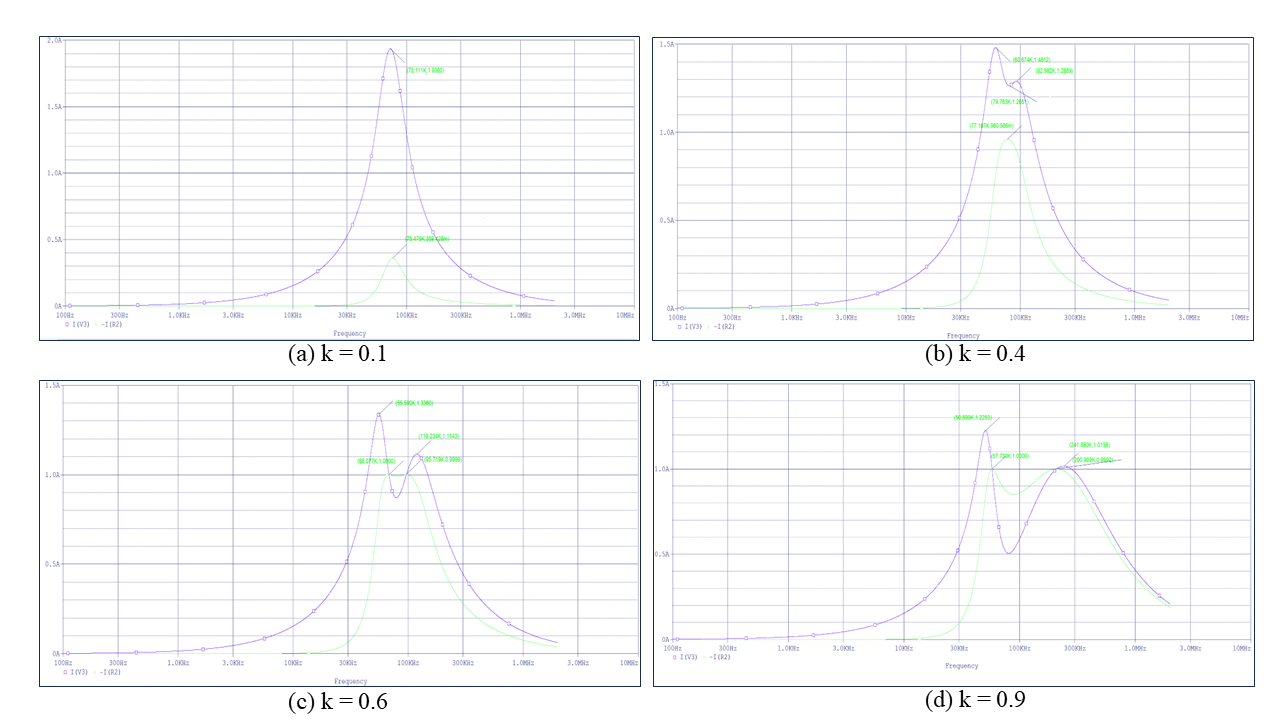}
    \caption{\textbf{AC Sweep result at various coupling coefficients. The purple line indicates the primary side current and the Green line indicates the secondary side current.}}
    \label{fig:AC Sweep}
\end{figure}

The purple graph is the primary side current and the Green graph is the secondary side current. At the above AC Sweep simulation, when the coupling coefficient is lowest(0.1) as described in (a), the primary side has maximum power at 72.11khz and the secondary side has maximum power at 75.475khz. Here, we can observe that the maximum power at each RX and TX side is shown at different frequency. So, it is a wrong approach to consider the maximum power transfer point at the TX side as the maximum power transfer point at the RX side, and needs to estimate RX side power. Also, we can discover that even before the frequency splitting does not occur at the secondary side current(green one), the frequency of the secondary’s single pole keeps increasing as k increases. From (b) where k=0.4, the frequency splitting Phenomenon occurs at the primary side, so the primary side power has a dual local max point. From (c) where k=0.6, the secondary side also shows frequency splitting. As two coils couple more strongly, the splitting is intensified, so the local max point moves further from the original resonant frequency. On the primary side, poles of lower frequency have bigger power than another one, but on the secondary side, both poles have the same magnitude of power at both frequency.

\subsection{System modeling by MATLAB / SIMULINK}
\begin{figure}[ht]
    \centering
    \includegraphics[width=.5\textwidth]{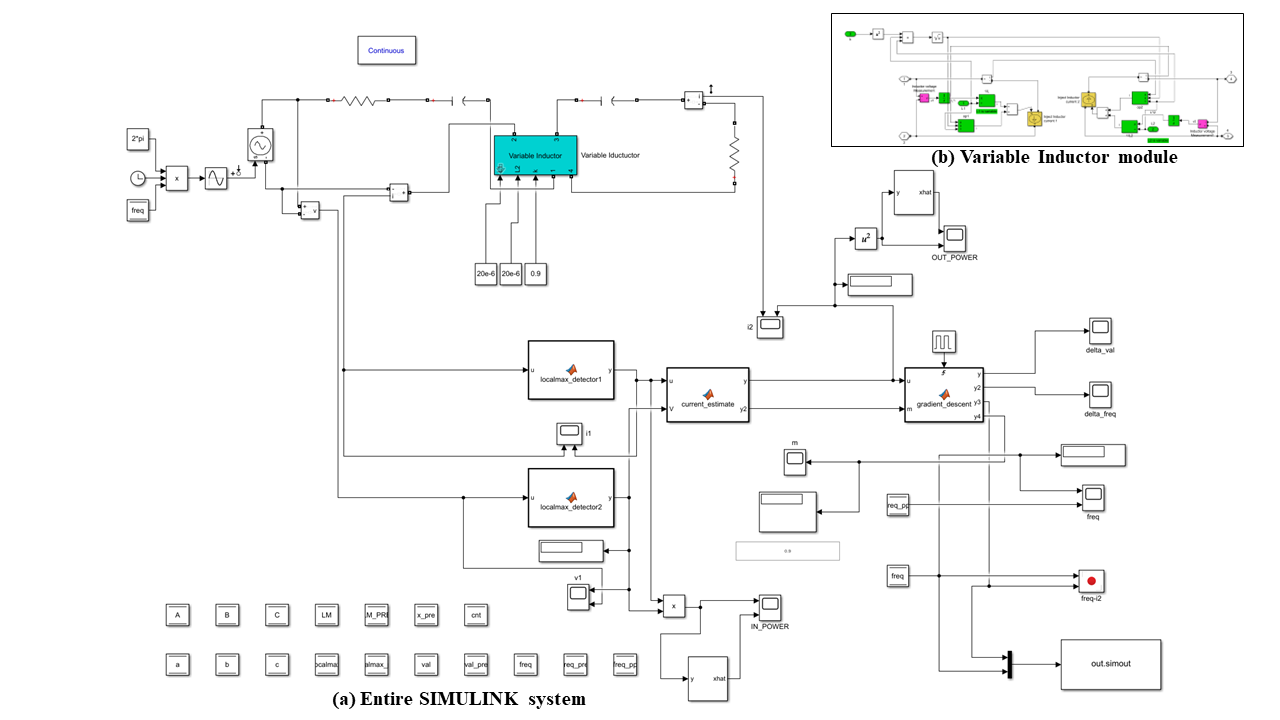}
    \caption{\textbf{(a) Adaptive Wireless Power Transfer system(WPT) implemented in SIMULINK; (b) Variable M mutual inductor block.}}
    \label{fig: Simulink system}
\end{figure}
The entire system can be modeled like Figure 5 in SIMULINK. SIMULINK doesn't support variable M Mutual inductor block, so we made customized variable M Mutual inductor block as shown in the inserted image (b) to change the mutual inductance gradually during transient simulation. The above SIMULINK circuit can be simplified more to the block diagram in Figure 6 for understanding. The blue block corresponds to the SIMULINK function block. All element’s value is same with the prior SPICE simulation.
\begin{figure}[ht]
    \centering
    \includegraphics[width=.5\textwidth]{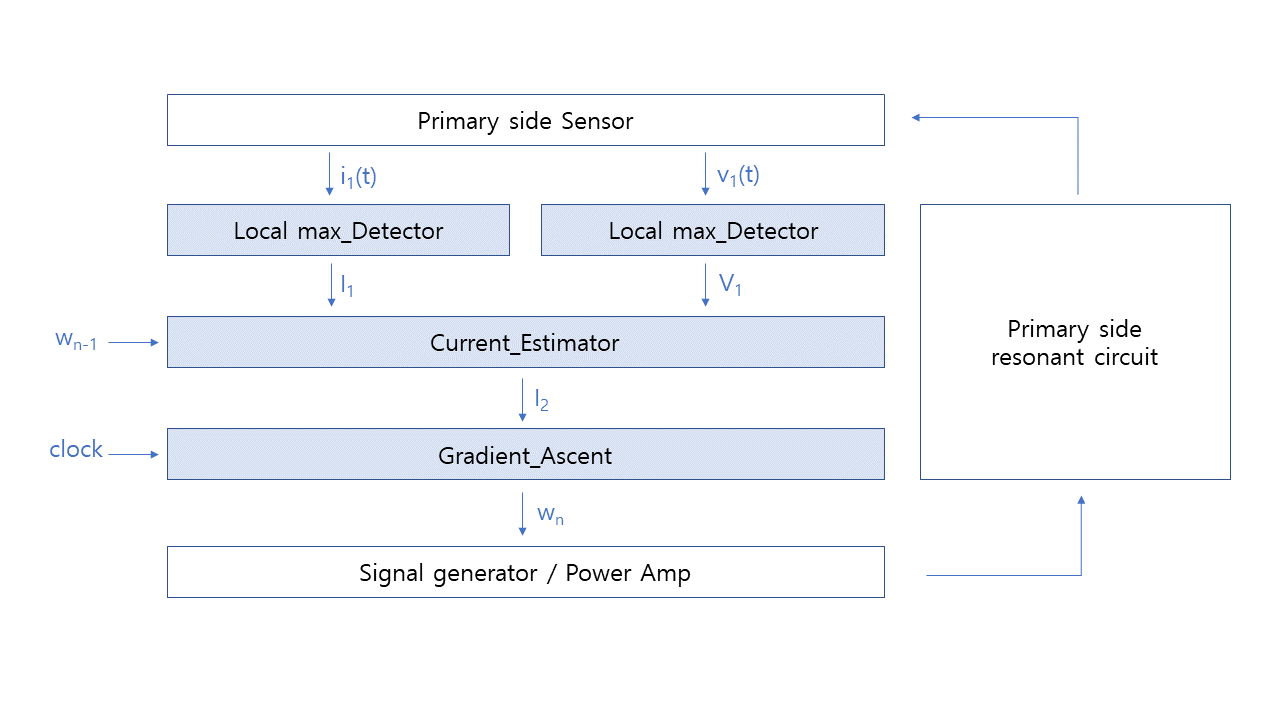}
    \caption{\textbf{Block Diagram of System}}
    \label{fig:Block Diagram of System Flow}
\end{figure}

The primary side resonant circuit is derived by an ideal voltage source which generates a sine wave of corresponding operating frequency. ‘Local max\_Detector’ block is used for detecting the amplitude of primary side current and voltage. ‘Current\_Estimator’ block estimates the secondary side current amplitude using the result from ‘Local max\_Detector’ and current frequency. ‘Gradient\_Ascent’ block executes the gradient descent algorithm on the period of T which is the period of the clock as described in Figure 6. The calculated frequency for the next step is fed into the signal generator and drives the primary side resonant circuit at that frequency. 
The operation of ‘Gradient Ascent’ block can be summarized in the pseudo algorithm of Table 2.

\begin{table}[ht]
\resizebox{.5\textwidth}{!}
{
\begin{tabular}{|l|}
\hline
\textbf{Initialization}: Initialize all variable to start value (usually zero)and drive system at   staring frequency \\ \hline
\textbf{while1}: Repeated every   $\delta$T                                                                              \\ \hline
Execute ‘Local max\_Detector’                                                                                  \\ \hline
Execute ‘Current\_Estimator’: keep estimating |$I_{2}$|   which is correlated to power                         \\ \hline
\textbf{while2}: Repeated every   $\Delta$T                                                                              \\ \hline
current = |$I_{2}$|                                                                                            \\ \hline
delta\_current = current– current\_past                                                                        \\ \hline
freq = freq\_past + 800000*delta\_current*((freq\_past-freq\_pp)/(abs(freq\_past-freq\_pp)))                   \\ \hline
freq\_pp = freq\_past                                                                                          \\ \hline
freq\_past = freq                                                                                              \\ \hline
current\_past = current                                                                                        \\ \hline
\end{tabular}
}
\caption{\textbf{Pseudo Algorithm of Adaptive WPT system}}
\label{Table:Table}
\end{table}
Again, the absorbed power at RX side is proportional to |$I_{2}$|, so we can use |$I_{2}$| to maximize RX power. The total simulation time is 0.025s and $\Delta$T which is the period of ‘Gradient Ascent’ execution is set to 0.00015s. If this value is large, the update of operating frequency gets low, so it takes a long time for the frequency to converge to local maxima. On the other hand, if this value is too small, the converge rate gets faster, but the update of frequency would be made before the current is stabilized, so this updated frequency might be made based on the wrong value, which causes the unstable of the entire system. The starting frequency is initialized to 75Khz and all other value is set to zero. The coupling coefficient is increased gradually from 0.2 to 0.9 by 0.1 during 0.02s transient simulation time.

\section{Result}
\begin{figure}[ht]
    \centering
    \includegraphics[width=.5\textwidth]{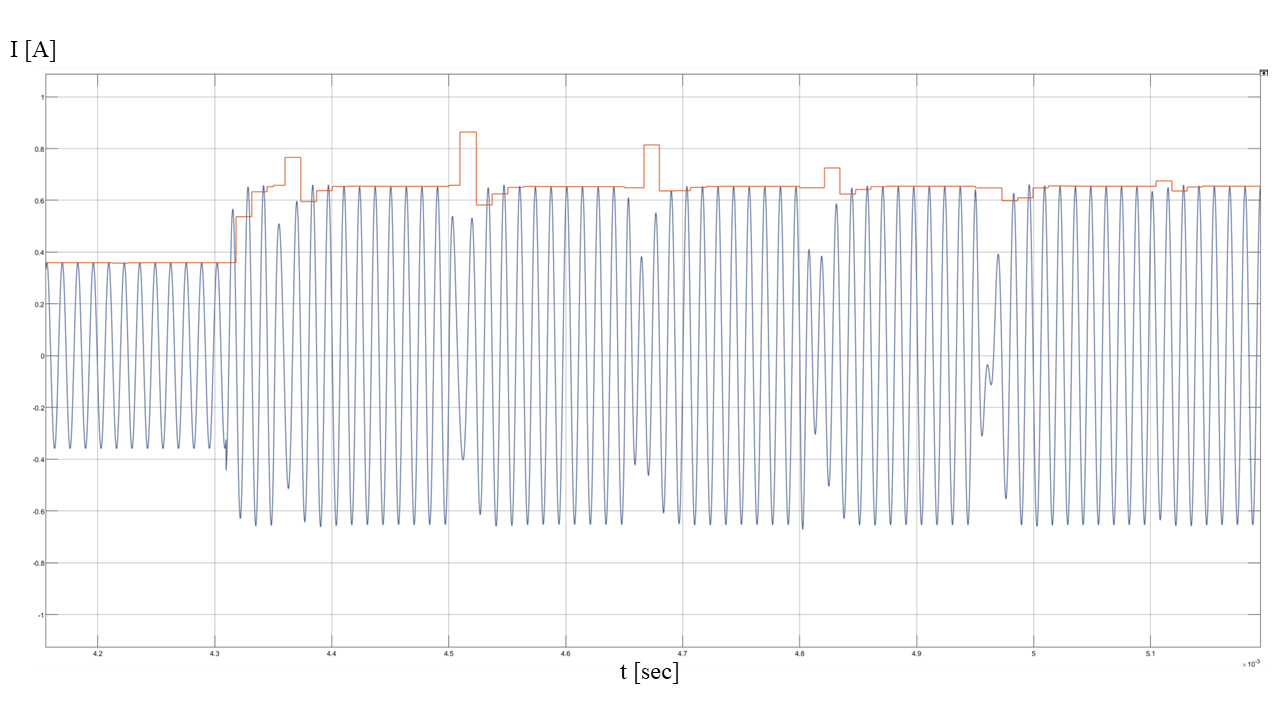}
    \caption{\textbf{Estimated secondary current amplitude(Red line) versus Actual secondary current waveform(Blue line)}}
    \label{fig:current}
\end{figure}

\begin{figure}[ht]
    \centering
    \includegraphics[width=.5\textwidth]{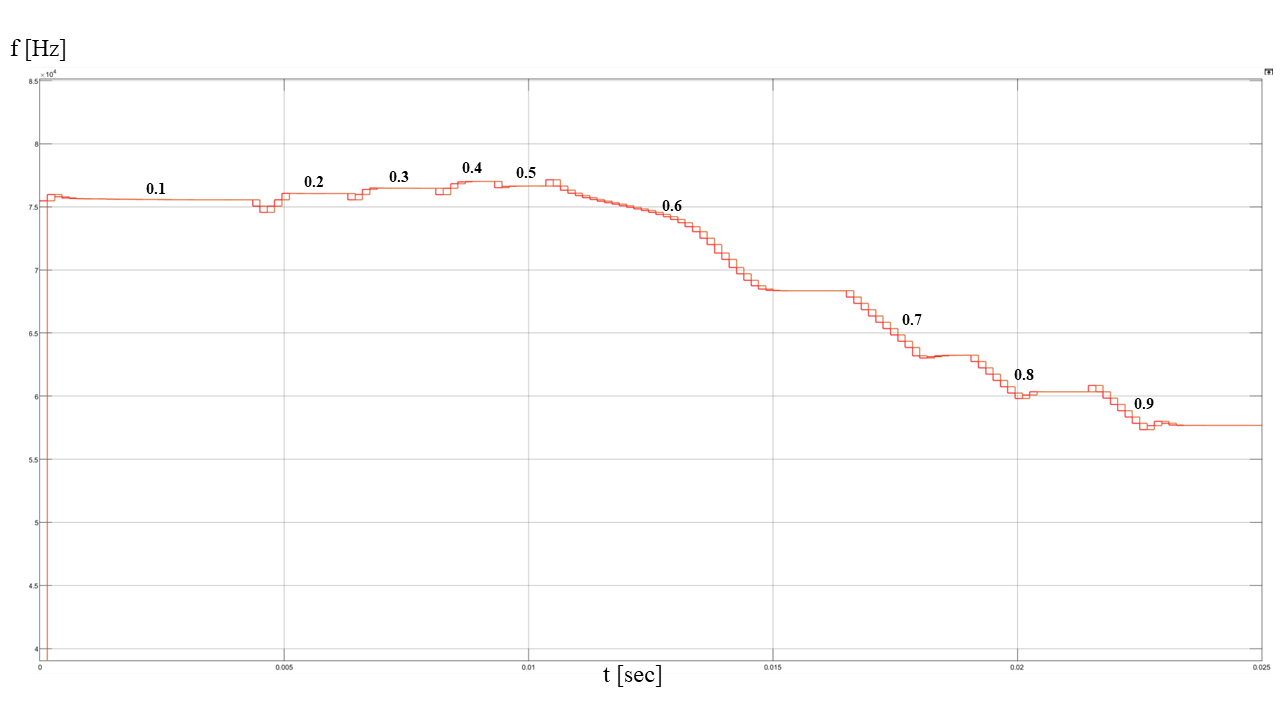}
    \caption{\textbf{Operating frequency at varying k(coupling coefficient) on transient simulation.}}
    \label{fig:freq}
\end{figure}

\begin{figure}[ht]
    \centering
    \includegraphics[width=.5\textwidth]{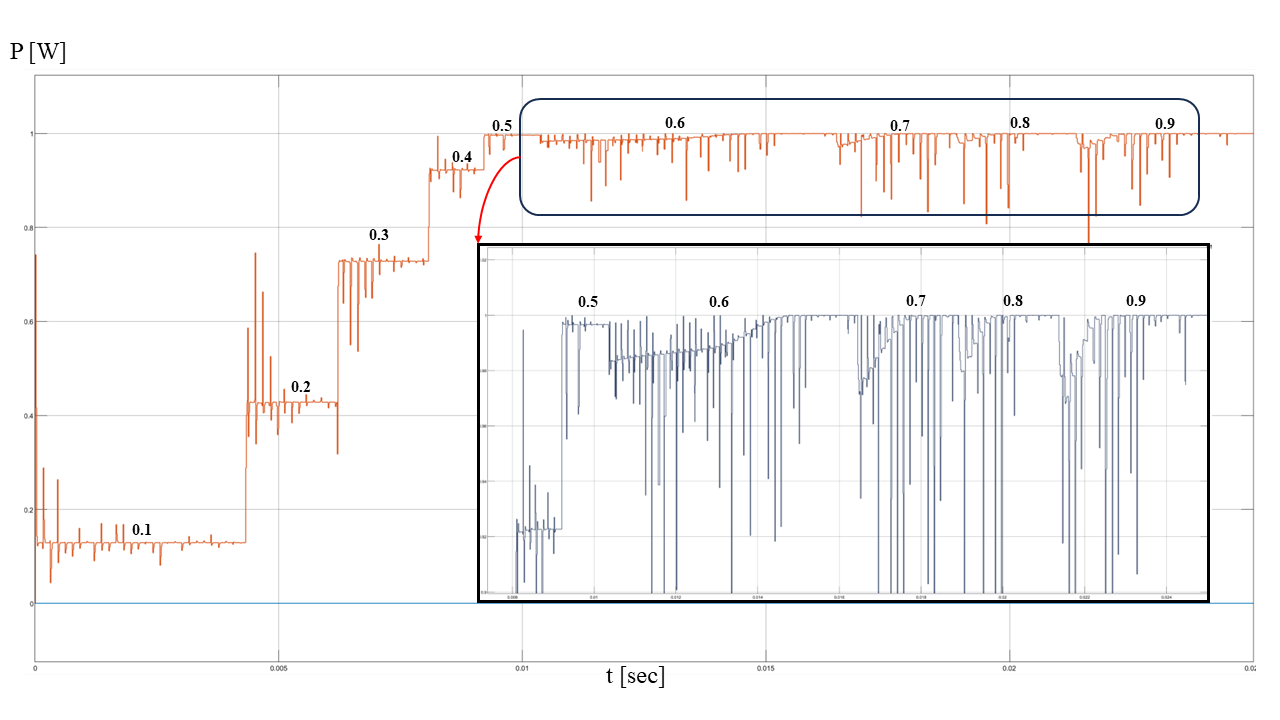}
    \caption{\textbf{Power consumption of RX side at varying k(coupling coefficient) on transient simulation.}}
    \label{fig:power}
\end{figure} 

As shown in Figure 7, the blue graph is a measured current waveform at secondary side and the red one is the estimated current amplitude using the primary side input impedance. After the system is stabilized enough after a frequency transition which is occur at every 0.00015s, the estimated current amplitude converges to the actual secondary side current amplitude and the error is almost negligible. So we can successfully confirm the validity of the equation for |$I_{2}$| and M.
Figure 8 is about the operating frequency f according to the variation of coupling coefficient k. We can observe that the operating frequency is dynamically changing to the maximum power transfer point as the coupling coefficient gradually increases from 0.1 to 0.9. From the point of k=0.6, frequency splitting begins to appear obviously, making the operating frequency drop-down largely. This result coincides with the SPICE simulation where the pole splitting of the secondary side current occurs from k=0.6. It can be seen that the operating frequency converges to the lower pole rather than the higher one when frequency splitting begins. This is because the gradient in the direction to the lower pole is greater when the frequency split happens.
In the case of delivered power shown in Figure 9, as the coupling gets strong, RX side power increases. After frequency splitting occurs at around k=0.6, RX power drops quite largely but as the operating frequency keeps tracking the maximum power transfer point by gradient ascent algorithm, RX power value increases slowly and converges to maximum value again. At k=0.8, 0.9, the system achieves its possible maximum value either.
Figure 10 and Table 3 show the comparison of RX power at various coupling coefficient between conventional wireless power transfer system which use static impedance matching optimized to k=0.5 and the proposed system in this paper which track maximum power transfer frequency dynamically, through SIMULINK simulation. Prior to frequency splitting, the delivered power by the proposed system was slightly higher than that of a Fixed frequency system, but it is almost the same. This slight superiority is because the frequency of the maximum power transfer point keeps varying even before the frequency splitting. After the frequency occurs around k=0.6, transferred power by the conventional method drops sharply, while the proposed method is capable of remaining maximum power transfer, achieving up to 30\% improvement. 

\begin{figure}[!t]
    \centering
    \includegraphics[width=.5\textwidth]{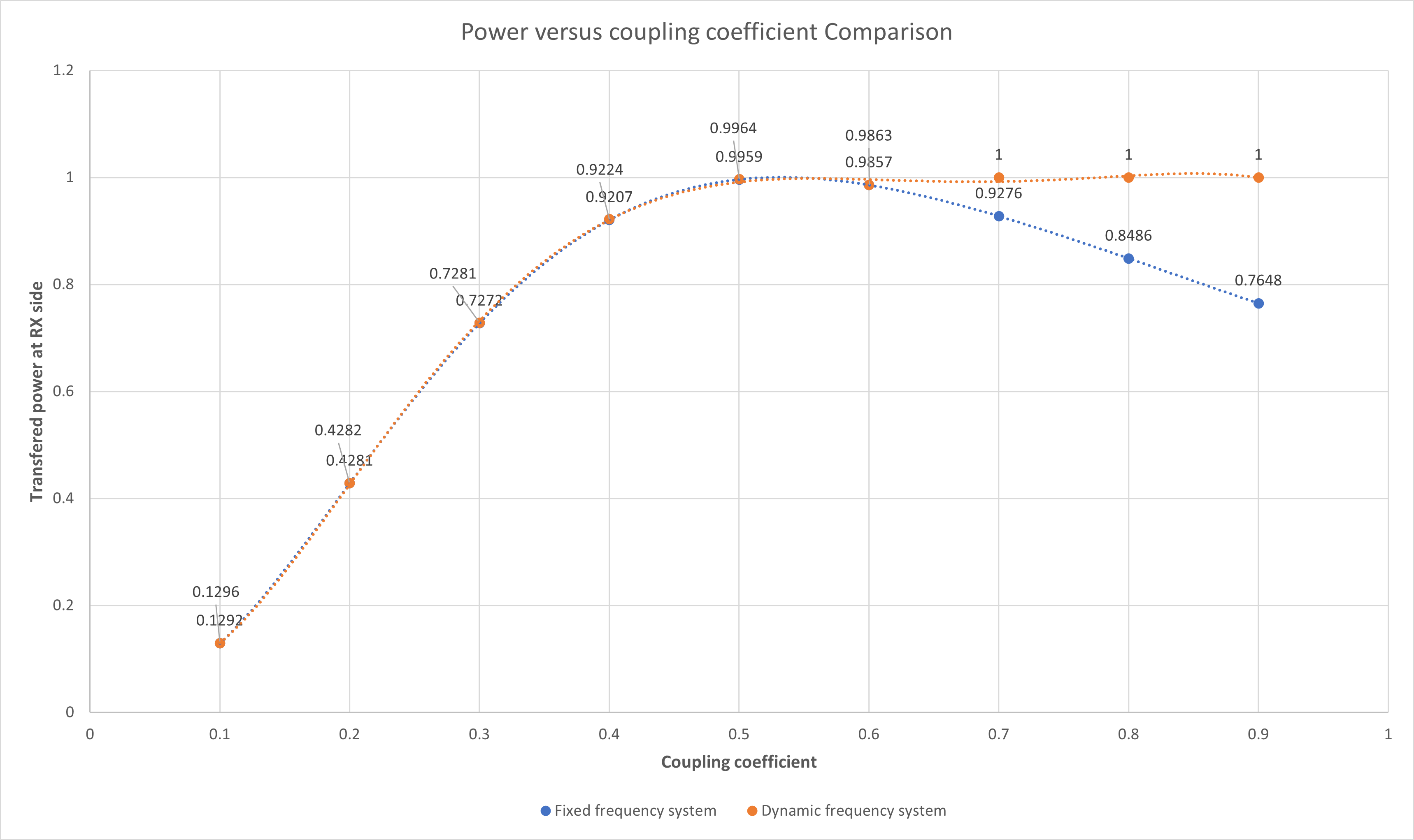}
    \caption{\textbf{Performance comparison between Static impedance matching and Adaptive matching scheme.}}
    \label{fig:result}
\end{figure} 
\begin{table}
\begin{tabular}{|l|c|c|c|}
\hline
\textit{\textbf{k}} & \multicolumn{1}{l|}{\textit{\textbf{Static matching [W]}}} & \multicolumn{1}{l|}{\textit{\textbf{Dynamic matching [W]}}} & \multicolumn{1}{l|}{\textit{\textbf{Improvement [\%]}}} \\ \hline
\textbf{0.1}        & 0.1292                                                   & 0.1296                                                    & 0.31                                               \\ \hline
\textbf{0.2}        & 0.4281                                                   & 0.4282                                                    & 0.02                                               \\ \hline
\textbf{0.3}        & 0.7272                                                   & 0.7281                                                    & 0.12                                               \\ \hline
\textbf{0.4}        & 0.9207                                                   & 0.9224                                                    & 0.18                                               \\ \hline
\textbf{0.5}        & 0.9959                                                   & 0.9964                                                    & 0.05                                               \\ \hline
\textbf{0.6}        & 0.9857                                                   & 0.9989                                                    & 1.33                                               \\ \hline
\textbf{0.7}        & 0.9276                                                   & 1                                                       & 7.80                                               \\ \hline
\textbf{0.8}        & 0.8486                                                   & 1                                                       & 17.84                                              \\ \hline
\textbf{0.9}        & \textit{0.7648}                                          & 1                                                       & 30.75                                              \\ \hline
\end{tabular}
\caption{\textbf{Improvement of prthe oposed method with respect to conventional static impedance matching method.}}
\label{Table:improvement}
\end{table}

\section{Conclusion}
In this research, the magnetic resonant coupled wireless power transfer scheme for the movable device is mathematically analyzed and modeled. We validated the above model by various simulations and confirmed that even without a communication link, the dynamic(adaptive) frequency tracking method can achieve maximum power transfer using only primary side information such as the magnitude of input impedance. Also, the proposed method is compared with the conventional static impedance matching method, and transfer performance is evaluated by simulation. In this way, the pros and cons of both methods were discussed.

\EOD

\end{document}